\documentclass[letterpaper]{article} 
\usepackage{aaai2026}  
\usepackage{times}  
\usepackage{helvet}  
\usepackage{courier}  
\usepackage[hyphens]{url}  
\usepackage{graphicx} 
\usepackage{rotating}
\usepackage{xcolor}
\usepackage{booktabs}
\usepackage{multirow}
\usepackage{graphicx}
\usepackage{natbib}  
\usepackage{caption} 
\usepackage{algorithm}
\usepackage{algorithmic}

\usepackage{newfloat}
\usepackage{listings}
\DeclareCaptionStyle{ruled}{labelfont=normalfont,labelsep=colon,strut=off} 
\floatstyle{ruled}
\newfloat{listing}{tb}{lst}{}
\floatname{listing}{Listing}

\renewcommand{\textsf}[1]{{\small\fontfamily{cmss}\selectfont#1}}

\title{When `For You' Isn't For You: \\Measuring User Agency in TikTok's Algorithmic Feed}
\author{
    Levi Kaplan, Devin Patel, Nicole Gerzon, Alan Mislove, Piotr Sapiezynski
}
\affiliations{
    Northeastern University, Boston MA\\
    \{kaplan.l, patel.devin1, gerzon.n, p.sapiezynski\}@northeastern.edu, amislove@ccs.neu.edu
}

\begin{document}

\maketitle

\begin{abstract}
The short-form video-sharing service TikTok has become an important platform in the social media landscape, with much of its popularity owed to its algorithmically-driven ``For You Page'' (FYP). This feature serves as the ``home screen'' for the platform and provides a personalized feed of content for each user. Unlike other social media services---where new users start their journey by explicitly signaling whom they choose to friend or follow---the TikTok FYP algorithm instead begins making inferences based on implicit signals, such as how long they watch particular videos. As a result, users have less explicit control over what content they see, and concerns have been raised about the impact on users (e.g., the delivery of potentially harmful content).

In this work, we investigate the extent to which users have control over the content they see on the FYP on TikTok. We first develop novel techniques to study the TikTok mobile app, introducing a new avenue for conducting controlled experiments that enable us to send both explicit and implicit signals on the app. We then use these techniques to study the FYP algorithm based on accounts we control. We find that the FYP algorithm is sensitive to both types of signals, changing the amount of personalized content the account sees. However, we find that users may have difficulty convincing the FYP algorithm to stop showing content the user wishes to no longer see:  the most effective explicit signal---marking a video as `Not Interested'---is unintuitively buried in the interface.  Worse, we find that once  accounts cease to indicate disinterest in a topic, many find their feeds dominated by such content again.
\end{abstract}

%

\section{Introduction}
TikTok is the most popular short-form video sharing service in the world~\cite{tiktok_global_most_pop}, with over one billion active users each month worldwide~\cite{tiktok_monthly_global_users}. Much of TikTok's popularity stems from its ``For You Page" (FYP), an algorithmically curated, personalized feed of videos that serves as the homepage for the platform~\cite{tiktok_fyp_popularity}. While an algorithmic feed of content is not unique to TikTok, their algorithm is said to be highly addictive due to how quickly and accurately it infers user interests~\cite{tiktok_read_mind}, even to the point of being called ``creepy''~\cite{tiktok_creepy}.  TikTok has given broad indications of how the FYP algorithm selects content, stating that it often uses {\em implicit} user signals (e.g., watching a video to the end) in addition to {\em explicit} user signals (e.g., ``liking'' a video).  Prior work found that following users (an explicit signal) and watch time (an implicit signal) had the greatest influence on the feed~\cite{boeker2022empirical}.  TikTok's ability to use implicit signals in the algorithm is underscored by their documentation, which states that ``the best way to curate your For You feed is to simply use and enjoy the app''~\cite{tiktok_how_fyp_control}. 

This focus on using implicit signals for recommendation raises concerns about user agency, or users' ability to control the content they are shown~\cite{wsj_2021}.  This has come up in discussions of TikTok's impact on younger users~\cite{tiktok_children_negative1}, creation of filter bubbles~\cite{filterbubble_reels_comaprison, filterbubble_samplett}, spread of misinformation~\cite{mask_misinformation}, and sharing of potentially harmful content~\cite{pryde2022tiktok, liu2021influence, lgbtqtiktokharms, weimann2023research}.  On platforms that rely heavily on explicit signals for personalization, users who are no longer interested in a certain type of content can simply unfollow or unfriend others who publish on that topic.  On platforms that rely heavily on implicit signals---such as TikTok---the actions that users should take, and the effectiveness of those actions, are less clear.  To provide user agency and control, TikTok has introduced various mechanisms for adjusting their feed, such as refreshing the For You page, using keyword filters to block specific content, and marking videos as `Not Interested'~\cite{tiktok_recommends_content}. However, the effectiveness of these tools and the extent to which they enable user control is not fully understood. 

In this paper, we focus on user agency in influencing TikTok's FYP algorithm. We aim to understand how quickly the algorithm personalizes content, and how effective the controls the platform provides to users are at ``de-personalizing'' it.  Unfortunately, studying TikTok presents a number of technical challenges.  Existing audits have primarily investigated TikTok through its web interface~\cite{boeker2022empirical, vombatkere2024tiktok, zeng2023okboomer}. Around two thirds of traffic to TikTok comes from mobile devices~\cite{semrush}, and the website has until very recently lacked many features present in the mobile app~\cite{tiktok_different_web}. Additionally, to the best of our knowledge, it is unknown whether the TikTok's behavior is the same across web and mobile devices, underscoring the importance of investigating the method chosen by most users.

To close this gap, we first develop infrastructure for instrumenting the mobile app and collecting the data it exchanges with the TikTok servers.
We further introduce a novel methodology of sock-puppet accounts called ``cloning'', using which we can test counter-factual scenarios on accounts with identical platform-use histories. This enables us to ask ``what if?'' questions and better understand how each action influences the FYP algorithm.  

We apply these techniques to study user agency in controlling the FYP algorithm, focusing on content with three topics: \textsf{cooking}, \textsf{fitness}, and \textsf{sports betting}. We find that both implicit and explicit negative signals are effective at reducing the amount of topic content delivered to the FYP, and that most of the time (but not always), explicit signals (e.g., marking a video as `Not Interested') are more effective at reducing unwanted content than implicit signals (e.g., skipping a video). 
Finally, we find that the FYP algorithm can often ``relapse'': many accounts which cease to express disinterest and begin watching topical videos again can see their feeds dominated by such videos.  This behavior is more common for \textsf{cooking} and \textsf{fitness} than for \textsf{sports betting}. 

Overall, our paper makes three contributions:
\begin{enumerate}
    \item{\em Novel auditing techniques.} Previous audits on TikTok have primarily focused on the TikTok web interface, which has lacked key features of the mobile app and is used by a minority of the TikTok user base. We present a comprehensive set of techniques for auditing the mobile app through emulator control, mobile app modification, and network traffic interception, providing researchers with the ability to run numerous controlled experiments. 

    \item{\em Novel techniques for cloning accounts.} We introduce a novel auditing technique that involves simulating network traffic sent to the TikTok algorithm to precisely duplicate the history of a given account.  We demonstrate that the duplicated accounts are treated indistinguishably by the FYP algorithm, enabling ``what if?'' questions.

    \item{\em Studying user agency in the FYP algorithm.} Using the novel methodologies above, we conduct interventional studies to understand the exact effect that user controls have on the resulting FYP. 
    We show that explicit signaling is often more effective than implicit signaling, and that in many cases, if a user re-engages with content they previously signaled disinterest in, the platform reverts to consistently showing that content.
\end{enumerate}

The remainder of this paper is organized as follows.  We provide background and overview related work in Section~\ref{sec:background}, introduce our methodology in Section~\ref{sec:methodology}, and detail our results in Section~\ref{sec:results}.  A concluding discussion is in Section~\ref{sec:discussion}.

\section{Background and Related Work}\label{sec:background}
We now provide background and describe related work.

\subsection{TikTok and Prior Audits}
Prior work has highlighted the importance of algorithmic audits on recommendation systems, as they have a serious impact on the information users can access~\cite{bandy2021problematic}. The FYP algorithm relies on user activity~\cite{tiktok_how_fyp_control} and therefore requires historical and real-time data to function. 

There are a number of different methodological approaches for conducting algorithmic audits~\cite{sandvig2014auditing}, of which at least two have been applied to TikTok: non-invasive user studies~\cite{zannettou2023leveraging, vombatkere2024tiktok}, and sock-puppet audits~\cite{boeker2022empirical, vombatkere2024tiktok, wsj_2021}.
These have focused on both the FYP algorithm and how it categorizes and pushes content~\cite{klug2021trick,scanlon2021app}, as well as the behavior of TikTok users and their subsequent outcomes~\cite{boeker2022empirical, vombatkere2024tiktok, wsj_2021, zannettou2023leveraging}. 

This paper provides a new perspective on the former using a sock-puppet audit framework~\cite{sandvig2014auditing}. Unlike previous studies that mainly focused on TikTok's web app \cite{boeker2022empirical, vombatkere2024tiktok}, we develop a novel methodology that can study TikTok's mobile application. Our research further distinguishes itself from prior work by implementing account cloning, which allows for duplicating sock-puppets from the same initial run and offers insight into different app feed control options.

In addition to more formal controlled audits, there is a substantial body of work that explores TikTok algorithmic ``folk theories'' through qualitative studies of user experience~\cite{karizat2021algorithmic, harris_tt_bnot}. Folk theories are socially-informed explanations of how an algorithm works in practice, believed to be true by users of the platform~\cite{folk_theories_historic}. Among them, there is scrutiny around whether TikTok's ``Not Interested'' feature works as intended, with users reporting that TikTok is ineffective in stopping the FYP algorithm from recommending more of the same, potentially damaging, content \cite{devito2022transfeminine, harris_tt_bnot}. 
We note that even if a formal audit can detect a change caused by the control features, it does not mean that change is commensurate with users' expectations.

\subsection{Studies on User Agency}
The concept of user agency refers to a user's ability to influence the content they see on a platform. TikTok's FYP is curated by an automated algorithm based on both explicit and implicit signals, such as watching, liking, and sharing recommended content~\cite{boeker2022empirical, kang2022tiktokuseragency, wsj_2021}. Similar recommendation-based platforms, such as YouTube's video recommendations or Facebook's ad delivery algorithm, have historically made it difficult for users to adjust inferred preferences~\cite{youtuberegrets, algorithmstrauma, algorithmstrauma2}. Even if the features to limit unwanted content exist, issues with effectiveness and usability makes their adoption rare~\cite{gunawan2021comparative, gak2022distressing, habib2022identifying}. This lack of user agency coupled with increased algorithmic personalization has dangerous effects such as amplifying filter bubbles~\cite{rabbithole2022general}, spreading misinformation~\cite{vaccineRabbithole}, and worsening existing harms~\cite{gak2022distressing}.  

\subsection{Fitness Content}\label{subsec:fitness}
With a growing number of people turning to social media platforms for advice on exercise and dieting, ``Fitness \& Gym'' ranks as one of the top categories for content and influencers across social media platforms~\cite{rogers2022communication, stollfuss2020communitainment}. While viewing content categorized as ``fitness related'', viewers are often exposed to content that features excessive dieting, body-shaming, and the glorification of eating disorders~\cite{pryde2022tiktok, liu2021influence}. Previous research has established that exposure to harmful fitness or dieting content, often demonstrated through popular tags such as ``\#thinspo'' and ``\#fitspiration'', can lead to increased negative body image and dissatisfaction~\cite{norton2017fitspiration, jeronimo2022effects}.  

\subsection{Sports Betting Content}\label{subsec:sportsbetting}
In 2018, the U.S.~Supreme Court ruled that states could legalize sports gambling.\footnote{Murphy v. National Collegiate Athletic Association, No. 16-476, 584 U.S. 453 (2018) [138 S. Ct. 1461]} Since then, the majority of U.S.~states have approved legalization and the sports betting industry now generates over \$1B in profit~\cite{milkedSportBetting}. Researchers have studied the harms of sports betting, which has been linked to gambling disorder and addiction~\cite{gainsbury2015internet}, with negative consequences seen in a higher proportion of younger users~\cite{mestre2022sports}.  Additionally, sports betting has the potential to be a far more challenging addiction to overcome. While traditional gambling required physical presence in a casino, sports betting is carried out completely virtually, which contributes to higher rates of addiction~\cite{milkedSportBetting, gainsbury2015online, hing2023situational}, and minimizes the negative feelings of loss resulting from unsuccessful bets~\cite{cantstop}.  Prior work has observed that sports betting content is prevalent on TikTok and is actively recommended by the FYP algorithm~\cite{ttsportsincrease}. 

\section{Methodology}\label{sec:methodology}

We now describe the methodology for running our audit on TikTok to understand user agency in the FYP algorithm. 

\subsection{Experimental Design}\label{subsec:design}
We aim to understand the effectiveness of the tools that TikTok provides in controlling the FYP algorithm. There are a number of mechanisms for users to provide signals to TikTok, including both implicit signals (e.g., a user scrolling past the video, or watching part of the video) or explicit signals (e.g., by clicking on `Not Interested').\footnote{We note that the `Not Interested' button is accessible rather nonintuitively via the `Share' menu or by long-pressing a video, and may not be known to all users.}  We study both types of signals. 

\begin{figure}
    \centering
    \includegraphics[width=.88\linewidth]{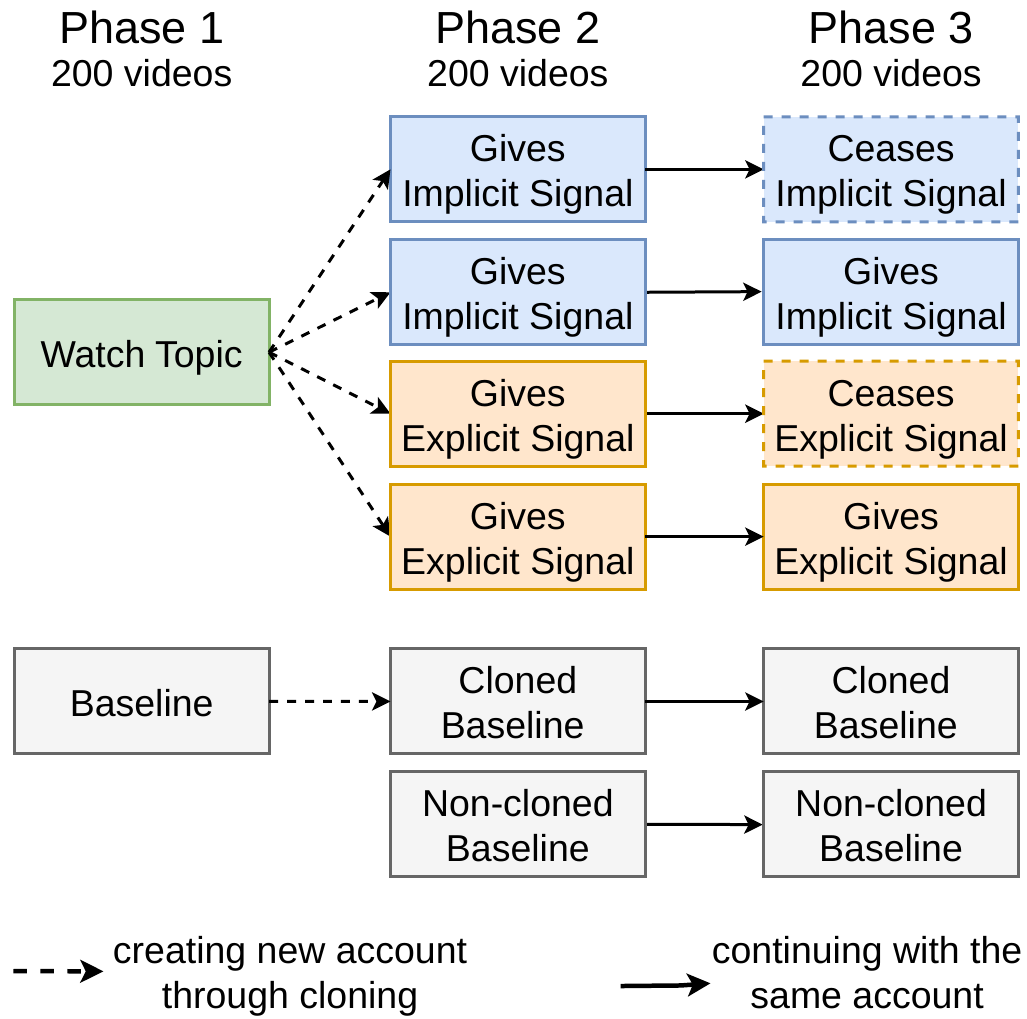}
    \caption{Steps for investigating user agency on TikTok. Each `Phase' occurs sequentially and involves scraping 200 videos from the FYP. Each element in the flowchart represents a different sock-puppet account ran on a different device. Devices in each column are run simultaneously.}
    \label{fig:ua-flowchart}
\end{figure}

Our experiments focus on answering three questions concerning user agency and the FYP algorithm:
\begin{enumerate}
    \item To what extent does the FYP algorithm personalize content based on user signals?
    \item To what extent does the FYP algorithm respond if the user signals they no longer wish to see certain content?
    \item To what extent does the FYP algorithm start showing content if a user---who previously signaled they did not want to see a particular type of content---stops doing so?
\end{enumerate}
We aim to answer these questions for three different types of content, which we refer to as the three `topics': \textsf{cooking}, \textsf{fitness}, and \textsf{sports betting}. For each topic, we run five separate experiments, each divided into three phases, as shown in Figure~\ref{fig:ua-flowchart}. The different experimental roles of the sock-puppet accounts are presented in Table~\ref{tab:device-type}.

We have intentionally chosen to run each Phase for the length of 200 videos, as it allows enough samples to provide statistical confidence for observed differences in algorithmic behavior without overwhelming TikTok's servers with requests or providing too much negative feedback so as to not receive topic videos in Phase 3.

\begin{table*}[t]
\renewcommand{\arraystretch}{1.5}
\begin{tabular}{@{}lll p{0.62\linewidth} @{}}
\textbf{Device}        & \textbf{Role} & \textbf{Phase} & \textbf{Description}                                                                                                                                                                        \\ \midrule
Watch Topic & Cloned                     & 1              & The account whose watch history is cloned by the treatment accounts. First undergoes a `seeding' process, where it watches 25 videos from the search feed, then scrolls through 200 videos from the FYP, watching all videos classified as related to the topic. Only the videos from the FYP are cloned.                                 \\
Baseline               & Cloned                     & 1              & The baseline account whose watch history is cloned by the Baseline Clone account. Does not undergo a seeding process, instead only scrolling through 200 videos without watching any videos.                                                   \\
Cloned Baseline         & Baseline                   & 2, 3           & Clone of Baseline. Represents a baseline that's watched the same number of videos as the treatment devices.                                                                                 \\
Non-cloned Baseline         & Baseline                   & 2, 3           & A fresh account that's not been cloned but runs simultaneously to the treatment devices. Represents a baseline that hasn't watched any videos.                                                \\
Gives Implicit Signal  & Treatment                  & 2, 3           & Clone of Watch Topic. Skips all videos regardless of whether they are classified as related to the topic.                                                                         \\
Gives Explicit Signal  & Treatment                  & 2, 3           & Clone of Watch Topic. Watches videos that are classified as related to the topic, then marks those videos as `Not Interested' afterwards.                                         \\
Ceases Implicit Signal & Treatment                  & 3              & Continuation of `Gives Implicit Signal'. Changes behavior of the Treatment, switching to again watching videos that are classified as related to the topic.  \\  
Ceases Explicit Signal & Treatment                  & 3              & Continuation of `Gives Explicit Signal'. Changes behavior of the Treatment, switching to again watching videos that are classified as related to the topic.  \\ 
\end{tabular}

\caption{Description of account types and their experimental roles.}
\label{tab:device-type}
\end{table*}

\subsubsection{Phase 1: Initial Personalization}
In Phase 1, we aim to understand to what extent the FYP algorithm personalizes content.  For a given topic, we first signal interest explicitly by searching for videos related to the topic and watching the first 25 videos from the resulting search results.  We then browse 200 videos from the FYP, and continue signaling interest implicitly by fully watching any videos that happen to correspond to the topic of the experiment (and skipping all other videos).\footnote{Whenever we say a video is skipped throughout the paper, we actually watch the video for a random amount of time between 0.2 and 2 seconds, similar to what a human user would do.}  We refer to this as the \textsf{Watch Topic} treatment.

Concurrently, we run a separate \textsf{Baseline} treatment, which does not undergo a similar seeding process and simply skips all videos. This allows us to compare the FYP videos shown to the \textsf{Watch Topic} treatment to what an account that did not signal such interest is shown.

\subsubsection{Phase 2: Signaling Disinterest}
In Phase 2, we aim to understand to what extent the FYP algorithm responds if users signal they no longer wish to see content on a given topic.  We test the efficacy of explicit signals (i.e., marking videos on the given topic as `Not Interested', named \textsf{Gives Explicit Signal}) and implicit signals (i.e., skipping videos on the given topic, named \textsf{Gives Implicit Signal}).

To create accounts for Phase 2, we clone\footnote{We only clone the videos seen in the FYP during Phase 1, and we do not clone the 25 videos watched from the search feed.} the accounts we created in Phase 1 (both the \textsf{Baseline} and the \textsf{Watch Topic}).
The cloning process is described in detail in Section \ref{sec:feed-cloning}.  For the \textsf{Watch Topic}, we actually clone it into four accounts:  two each for explicit and implicit signaling.  We do so as we will treat some of these accounts differently in Phase 3.  For the \textsf{Baseline}, we clone the account into a \textsf{Cloned Baseline} account for consistency in the experimental setup.  
Finally, we introduce a new account that has not seen any prior videos, which we call \textsf{Non-cloned Baseline}. This account behaves identically to the baseline account from Phase 1; the only difference is that it is not a clone and has not seen any videos beforehand.  We run all of these accounts concurrently for 200 videos.

For statistical significance, we repeat the experiment structure in Figure~\ref{fig:ua-flowchart} five times for each of our three topics, each with different accounts, for a total of fifteen runs. We then continue these runs in Phase~3. 

\subsubsection{Phase 3: Interest Relapse}
In Phase~3, we aim to see to what extent the FYP algorithm responds if a user stops signaling disinterest in the given topic, and instead starts watching such videos (as in Phase 1).  We aim here to understand whether TikTok offers opportunities for a user to `relapse', reverting to a previous behavior or interest which for a period of time they tried to curb.

For each of the two \textsf{Gives Explicit Signal} and \textsf{Gives Implicit Signal} accounts from Phase 2, we have one switch behavior and start watching videos on the topic; we name these \textsf{Ceases Explicit Signal}  and \textsf{Ceases Implicit Signal}.  For the other account for each, we have it continue giving explicit or implicit signals as in Phase 2.  We are interested in whether we detect a `relapse' effect, which we define as a statistical difference between the amount of topic content delivered to \textsf{Gives Implicit (Explicit) Signal} and \textsf{Ceases Implicit (Explicit) Signal}. We also continue running the \textsf{Cloned Baseline} and \textsf{Non-cloned Baseline} accounts.  We run all of these devices for 200 videos on the FYP feed.

\subsection{Auditing the TikTok Mobile App}\label{methodology_audit}
Implementing the experimental design described in the previous section requires overcoming a number of technical hurdles; we describe our approach below.

\subsubsection{Emulating the TikTok App}\label{subsec:emulate}
To run TikTok on devices that enable automatic control, we employ Android device emulators running through Android Studio. This allows us to simulate many devices at once and control them using our computer. We use the Android Studio CLI\footnote{https://developer.android.com/tools} to launch the devices, and Android Debug Bridge (\texttt{\small adb})\footnote{https://developer.android.com/tools/adb} to connect to our devices and install the TikTok app. We use UIAutomator2\footnote{https://github.com/appium/appium-uiautomator2-driver} to simulate physical interactions with the devices, in particular swiping and clicking.  

\subsubsection{Creating Accounts}
Prior to running an experiment, we create all necessary TikTok accounts (including separate accounts for any baseline and cloned treatments). We create six new accounts for each of our fifteen experiments. All account creation is done automatically and simultaneously per experiment. We first reset the Mobile Advertising ID\footnote{https://support.google.com/admanager/answer/6274238?hl=en} of our Android emulators to prevent any carry-over effects from TikTok storing information based on advertising ID. To further avoid cross-experiment effects, we also reset the data of our TikTok app. We then automatically create TikTok accounts, using randomly generated email addresses. We monitor and manually solve any Captchas that may come up, which are occasionally required to create a TikTok account. After creating an account, TikTok asks the user whether they want to create an account nickname and whether there are any topics they are interested in.  We press `Skip' for both. Finally, a pop-up appears asking whether we approve TikTok accessing our device's contact list for use in finding other users we may know, which we deny. 

\subsubsection{Intercepting Network Traffic}
To run our experiments, we need to collect information for each video shown in the FYP feed.  While some of the metadata for a video is shown on screen (such as truncated like, comment, and save counts), other metadata fields such as the ID of the video, the length of the video, and the play count of the video are not as easily accessible. However, this additional metadata is available in the HTML of the video's corresponding web URL. We could gather this data by clicking `Share' on every video and obtaining the URL~\cite{mousavi2024auditing}, but doing so may be interpreted as a signal of interest by the FYP algorithm.  Instead, we intercept the network traffic of the TikTok app, and develop techniques to parse it and find each video's URL. 

Intercepting the app's network traffic requires bypassing TikTok's TLS certificate pinning, which protects the network traffic to and from the device from being intercepted and read. We un-pin the certificate using an open-source script~\cite{tiktok_ssl_bypass}. We then use the Monster-in-the-Middle technique via \textit{Mitmproxy}\footnote{https://mitmproxy.org/} to intercept the network traffic and collect it.\footnote{We note that we are only intercepting the data that our emulated device sends and receives, and we are not collecting or modifying any other users' personal data.} We configure each Android emulator to send all traffic via a separate HTTP proxy, each with a Mitmproxy instance listening to and saving the network traffic.  We then parse this saved network data to find the URL of the video and retrieve the metadata from the HTML. 

\subsection{Deciding Whether to Watch Videos} \label{sec:watch_decision}
Now that we have the ability to control the TikTok app and retrieve all the necessary video metadata, we need to determine \textit{how} our accounts interact with TikTok. 

\subsubsection{Choosing Topics}
We chose three interest topics for our research into user agency: \textsf{cooking}, \textsf{fitness}, and \textsf{sports betting}.  These topics have different levels of prevalence and present different potential harm to TikTok users. \textsf{Cooking} is highly prevalent, and is likely fairly innocuous.  While a user could decide they do not want to see cooking videos for mental health or other reasons, the exposure to cooking content is unlikely to be particularly harmful.  \textsf{Fitness} is also highly prevalent, and could be considered potentially harmful by some (e.g., pushing unrealistic body standards~\cite{jeronimo2022effects,liu2021influence, norton2017fitspiration}).  Finally, \textsf{sports betting} is not a highly prevent topic, and is considered potentially harmful by some (e.g., by promoting speculative gambling and risky financial decisions~\cite{mestre2022sports, gainsbury2015internet}).  
In fact, TikTok has specific policies in their Community Guidelines that limits when content related to body image\footnote{https://www.tiktok.com/community-guidelines/en/mental-behavioral-health} and gambling\footnote{https://www.tiktok.com/community-guidelines/en/regulated-commercial-activities} can be shared.

\begin{table*}[t!]
\begin{tabular}{@{}l p{0.43\linewidth} ccccc@{}}
 & & \textbf{Human} & \multicolumn{4}{c}{\textbf{ChatGPT}} \\
\textbf{Target Topic} & \textbf{Keywords}                                                                                    & \textbf{Fleiss' Kappa} & \textbf{Acc.} & \textbf{Prec.} & \textbf{Recall} & \textbf{F1} \\ \midrule
Cooking               & cooking, recipes, viral recipes, cooking tips, baking     & 0.922                        & 0.943                 & 0.966                  & 0.791                   & 0.87                \\[6pt]
Fitness               & fitness, health, exercise                                                                         & 0.937                        & 0.940                  & 0.992                  & 0.885                   & 0.935               \\[6pt]
Sports Betting        & sports betting, parlay, fantasy sports, sports gambling  & 0.963                        & 0.956                 & 0.992                  & 0.915                   & 0.952               \\ 
\end{tabular}
\caption{Keywords used in the ChatGPT prompt, and related validation scores for classification. On the left, we present the Fleiss' Kappa agreements for our human raters (N=4). Given their high agreement, we compare the human labelers' majority vote to ChatGPT's classification and present the accuracy, precision, recall, and F1 score. We overall see very high scores, justifying the use of ChatGPT for classification.}
\label{tab:verify-chatgpt-scores}
\end{table*}

\subsubsection{Classifying Videos}\label{subsubsec:classify-videos}
Next, we need to determine whether a given TikTok video is on one of the three topics above.  Prior work~\cite{boeker2022empirical} has used hashtag matching for such classification, but we opt against this for a few reasons. There is necessarily a selection bias for these hashtags; we do not know all the hashtags that are used on the videos we are interested in watching for a given experiment, and new hashtags can come up frequently or use certain euphemisms or codes to get around potential hashtag suppression by the algorithm~\cite{censorship_steen2023you, tt_censorship_dawson2024you, cobb2017not}. Additionally, many videos do not have hashtags in their description, meaning we would not be able to determine whether to watch those videos. 

Instead, we use ChatGPT 3.5 Turbo,\footnote{https://platform.openai.com/docs/models/gpt-3-5-turbo} a Large Language Model developed by OpenAI, to perform our classification. Through their API,\footnote{https://openai.com/api/} we provide ChatGPT with the video's description, suggested words (a list of related terms that TikTok automatically generates for videos), and hashtags, as well as the nickname (non-unique user-chosen name) and signature (profile biography) of the user who uploaded the video, and ask ChatGPT whether the video is on topic.  We provide the exact prompt in the Appendix, Section~\ref{subsec:iterrater}.

To understand the accuracy of ChatGPT, we compare ChatGPT's classification to human raters.  Specifically, for each topic, we selected 150 on-topic videos by searching for each topic using keywords on TikTok and taking the top 150 results.  We also selected 150 random videos from the FYP feed.  We shuffled the videos, and gave four of the authors of this paper a binary classification labeling task, presenting the same video information that we provide to ChatGPT.  For each video, they were asked a yes/no question whether the video was on the given topic.

To measure inter-rater agreement, we use Fleiss' Kappa, which allows us to correct for random chance agreement when measuring inter-rater agreement across multiple raters~\cite{fleiss}. Table~\ref{tab:verify-chatgpt-scores} presents the topics, the keywords used in the prompt, and their corresponding inter-rater agreements.  We first examine the inter-rater agreement among the humans; we see that the Fleiss' Kappa values are 0.922, 0.937, and 0.963 for \textsf{cooking}, \textsf{fitness}, and \textsf{sports betting} respectively.  These values are all high, and are well-above the threshold of 0.81 presented by Landis and Koch for `almost perfect' agreement~\cite{landis_koch}. To measure the performance of ChatGPT, we take a majority vote from our four human raters, marking disagreement in the case of a tie. We then analyze the accuracy, precision, recall, and F1 score of the performance of ChatGPT compared to the majority vote. These values are presented in Table~\ref{tab:verify-chatgpt-scores}, for each topic. Overall we find very high values---the only value of potential concern is the recall for the \textsf{cooking} topic, at 0.791. 
A lower recall here indicates that it is not classifying some cooking videos as related to cooking. 
This would mean that we provide a lower bound for how personalized a feed is, as it likely contains more cooking content than we report.
Given the high precision and accuracy, we believe this approach is sufficient for our classification task.

\subsection{Cloning Accounts} \label{sec:feed-cloning}
Finally, we describe our methodology and validation of our approach to TikTok account cloning.
Our goal with cloning accounts is to create copies of an account that will share identical watch/skip history, so as to provide the personalization algorithm with the exact same input, resulting in similar personalization outcomes. Pairs of accounts created this way will then enable us to test the efficacy of different treatments. We note that we do not aim to replicate real human behavior, instead seeking to isolate the effects of negative signaling on the resulting algorithmically delivered feed. Additionally, cloning does not imply that they will be delivered identical feeds, due to the inherent randomness of the TikTok algorithm; only that their past histories will be identical and we could expect that they have similar degree and type of personalization in their feeds. 
To clone an account, we first record the network traffic that it sends, then alter it such that it looks like it came from the new account, and then replay the relevant content back to TikTok's undocumented APIs, without actually watching any videos from the new account.
As we show below, this allows us to effectively spoof TikTok into personalizing the FYP as if the new account actually watched the videos we indicated.

\subsubsection{Recording and Generating Traffic}
By analyzing the intercepted network traffic of the \textsf{Watch Content} account, we found that TikTok sends the watch behavior of an account through several API endpoints.\footnote{We found that it uses the \texttt{\small /aweme/v1/aweme/stats} \texttt{\small /tiktok/v1/realtime/feedback}, \texttt{\small /aweme/v2/feed}, and \texttt{\small /service/2/app\_log/} API endpoints.} While some of these API endpoints receive data about user behavior as either JSON or form data, the \texttt{\small /aweme/v2/feed} endpoint uses a complex protobuf request body.\footnote{Protobuf (short for `protocol buffers') is a data format used to serialize data in such a way that the resulting format is smaller than equivalent payloads in JSON or XML.} We performed a static analysis of the TikTok app and discovered that it used Square's Wire library\footnote{https://github.com/square/wire} to serialize the data, enabling us to generate request data for arbitrary videos.  We amalgamate this process of automatically decompiling Square Wire Java protobuf classes in an open-source and general-purpose decompiler.\footnote{https://github.com/lumaaaaaa/protoextract} 

Similarly, the \texttt{\small /service/2/app\_log/} endpoint receives data in the \texttt{\small application/octet-stream;tt-data=b} format. By reverse-engineering the relevant code responsible for generating these request bodies, we determined that the data sent to this endpoint is compressed data using Zstandard\footnote{https://github.com/facebook/zstd} with a custom compression dictionary. We pinpointed that this dictionary is loaded from an AES encrypted file at runtime, and extracted the relevant key and IV to decrypt the file ourselves. Re-implementing this compression process in our framework allows for compressed payloads to be generated directly. 
 
\subsubsection{Reproducing Security Headers} The ability to generate data payloads alone is insufficient to effectively clone user data on TikTok. This is due to TikTok's use of security headers, primarily concerning a proprietary hashing algorithm \texttt{\small ttEncrypt}. This algorithm hashes the entire HTTP request and appends these hashes in various headers. When the server receives a request, it computes the same hashes for that request and compares them server-side, rejecting the request if the hashes do not match. To properly include these security headers, we use Frida\footnote{https://frida.re/} to hook the \texttt{\small ttEncrypt} function at runtime.\footnote{Frida is a general purpose dynamic instrumentation framework that enables hooking into the function in real time, allowing us to export the call to the hashing algorithm directly to Python.} We can then use the Android device as a `zombie', hooking the \texttt{\small ttEncrypt}  function at runtime to sign the requests we generate in Python, appending valid signature headers to each request sent.

\begin{figure}[t!]
    \centering
    \includegraphics[width=1\linewidth]{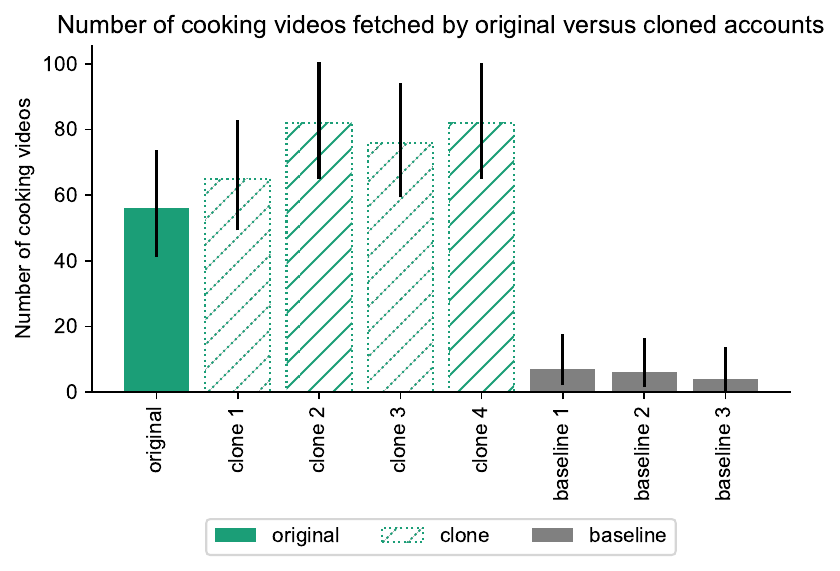}
    \caption{Our Account Cloning technique is successful at transferring personalization from the original account to the cloned account. Bars represent Agresti-Coull 99\% confidence intervals.}
    \label{fig:cloning-successful}
\end{figure}

\subsubsection{Verifying Cloning}
Before using the cloning methodology in our experiments, we first verify that the personalization of a cloned account is similar to the original. To do so, we perform a similar process to Phase 1 (Section~\ref{subsec:design}) but without first `seeding' the algorithm, instead watching videos from the default FYP that are related to cooking. This is because our cloning technique is only able to duplicate watching videos from the FYP and cannot replicate searching for videos and watching videos in the search feed. We refer to this account as the \textsf{original account}.  Using separate emulators, we then clone the watch history of this account four times; we refer to these as the \textsf{cloned accounts}. If cloning works, then we would expect the subsequent FYP feed delivered to the original account to be similar to those delivered to the cloned account.  For reference, we also include three separate \textsf{baseline} accounts with no history and which are not cloned, to measure the fraction of cooking videos that we can expect would appear in the FYP feed randomly.

If we were to take these eight accounts and scrape the FYP feed, we would necessarily be biasing the algorithm. This is because any amount of watch time acts as a signal to the algorithm: if we simply skipped through a set of videos, we would see fewer topic videos over time, and if we instead watched the videos, we would see more topic videos over time. Therefore, to understand whether our cloning methodology works without affecting the algorithm, we make use of another API hidden in the TikTok app. Whenever the FYP feed is fetched by an account, the \texttt{\small/api/v2/feed} endpoint is called. We discovered we can call this endpoint successively and it returns the next $N$ videos from the feed \textit{without} biasing the algorithm via sending whether the video was watched or not. We call this process \textit{fetching} the feed. By fetching the next 200 videos, we can get an idea of the level of personalization in the algorithm at a given moment, without biasing the algorithm by sending negative or positive watch signals from scrolling through the feed.

For all eight accounts, we fetch the next 200 videos from the FYP endpoint using this process. We hypothesize that the distribution of delivery of cooking videos to the original and cloned accounts' feeds will be similar, and that they will be different from the baseline. The results of this experiment can be found in Figure \ref{fig:cloning-successful}, with bars representing the Agresti-Coull 99\% confidence intervals. We see that all four clones see a considerably higher quantity of cooking videos than the baselines, and that the clones are not statistically distinguishable from themselves or from the original account, indicating that the personalization from the original device successfully transferred over.

\begin{figure}[t!]
    \centering
    \includegraphics[width=1\linewidth]{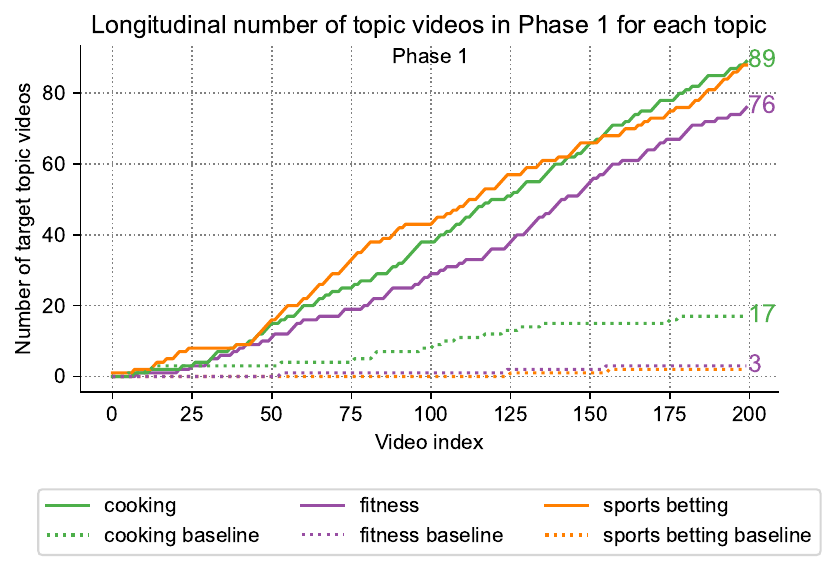}
    \caption{Personalization levels for the three different target topics: \textsf{cooking}, \textsf{fitness}, and \textsf{sports betting}. We see a similar degree of personalization over time between our topics.}
    \label{fig:phase-1-results}
\end{figure}

\begin{figure*}
    \centering
    \includegraphics[width=1\linewidth]{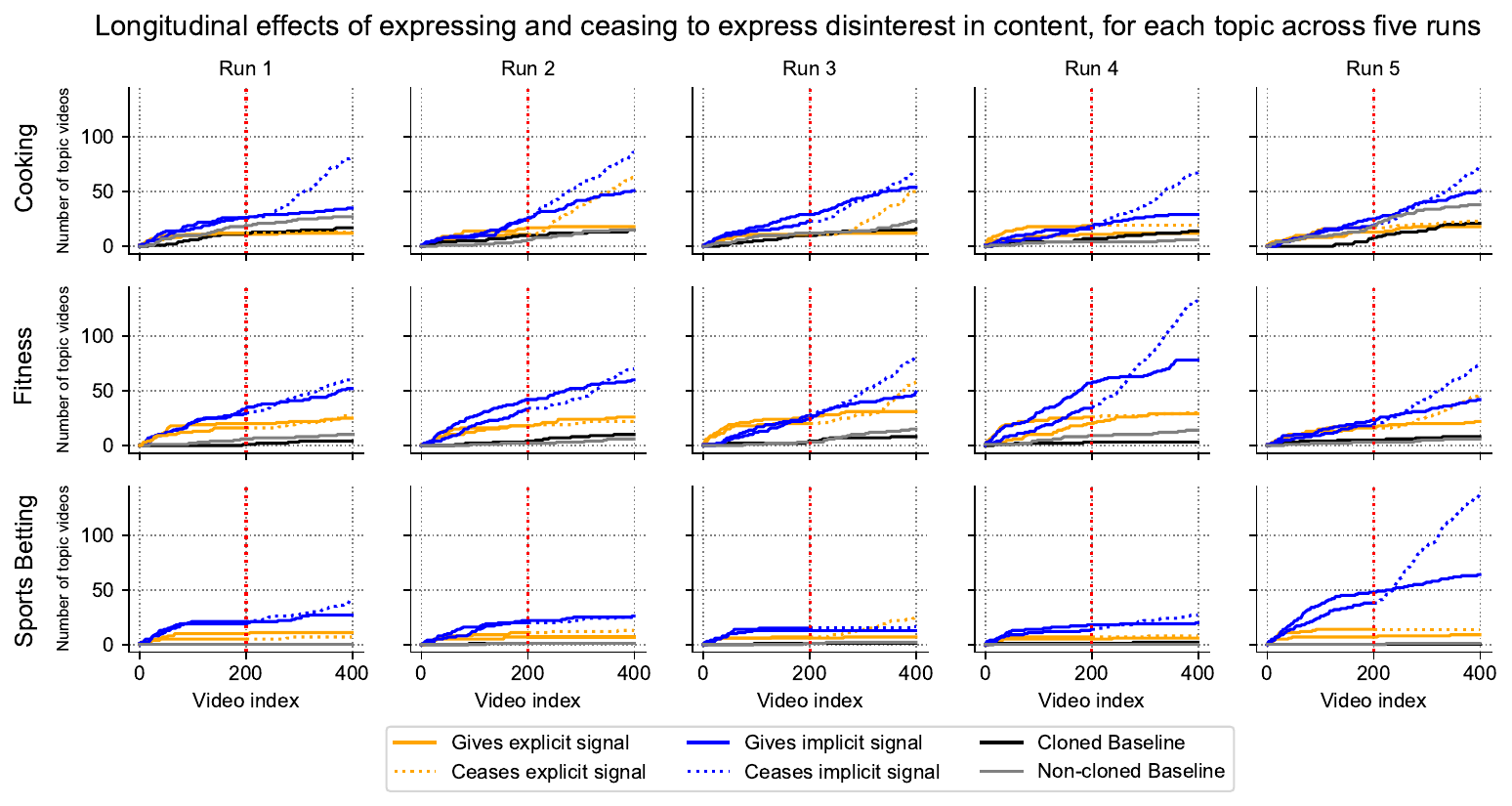}
    \caption{The number of topic videos delivered to our accounts across our fifteen experiments. The results of Phase 2 can be found on the left side of the red dotted line, and the results of Phase 3 can be found to the right side of the red dotted line, for each experiment. Dotted blue (orange) lines represent accounts which cease implicit (explicit) signaling, going back to watching topic content.}
    \label{fig:all-fifteen-runs}
\end{figure*}

\section{Results}\label{sec:results}

We now present the results from our study. 

\subsection{Phase 1: Initial Personalization}\label{subsec:phase1}
Recall that our goal in Phase 1 is to understand to what extent the FYP algorithm personalizes content.  For each of the three topics, we first `seed' an account by searching for the topic and watching the first 25 videos, and then collect 200 videos on the FYP feed (watching any on-topic videos and skipping all others).  We also run a separate baseline account at the same time which skips all videos.

Figure \ref{fig:phase-1-results} shows the number of topic videos delivered over the course of the experiment (after the `seeding') for each topic.  The $x$-axis represents the index of the video in the FYP feed (from 1 to 200), and the $y$-axis indicates the number of videos on the topic in the FYP feed at that point. At $x$=200, the $y$ value therefore represents the total number of target topic videos seen by the device over the phase.  

We see that for all topics, the algorithm quickly personalizes the content delivered via the FYP algorithm.  We achieved a similar degree of personalization for all three topics: the \textsf{cooking} and \textsf{sports betting} topics saw 89 (44.5\%) videos related to their respective topics. The \textsf{fitness} topic saw sightly fewer at 76 (38\%). Looking at the topics' prevalence in the baseline, we see 17 videos related to cooking (8.5\%), and only three related to fitness and sports betting (1.5\%).

\subsection{Phase 2: Signaling Disinterest}
\begin{figure}[b!]
    \centering
    \includegraphics[width=1\linewidth]{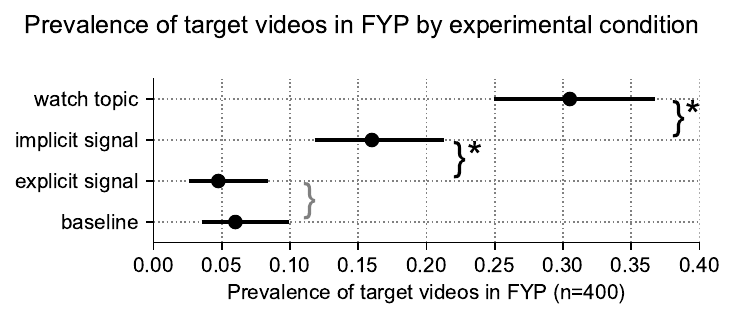}
    \caption{The account which sends an implicit positive signal (watch topic) receives more such videos than the account which sends an implicit negative signal. Sending an explicit negative signal results in even fewer on-topic videos, bringing the FYP close to the non-personalized baseline. The $\star$ symbol in the figure marks which differences in prevalence are significant at the 99\% confidence level.}
    \label{fig:watch-versus-signaling}
\end{figure}

In Phase 2, we aim to understand the extent the FYP algorithm responds when a user signals---either implicitly or explicitly---that they no longer wish to see the content that they previously engaged with. To that aim, we first run a simple experiment to see whether negative signals impact the FYP algorithm's recommendations at all. We focus on \textsf{cooking}, using six devices that have been cloned from a Phase 1 cooking account: two devices that watch all cooking videos, two that signal disinterest implicitly, and two that do so explicitly. 
We also include two baseline accounts.  
To measure statistical significance across the two accounts of each treatment, we add up the total number of on-topic videos seen as the sample proportion and use 400 (200 videos for each account) as the sample size. We then run a two-proportion Z-test at the 99\% confidence level. Figure~\ref{fig:watch-versus-signaling} shows that there were fewer target videos in the feeds of accounts who changed their behaviors, compared to those that did not.

\begin{figure*}
    \centering
    \includegraphics[width=1\textwidth]{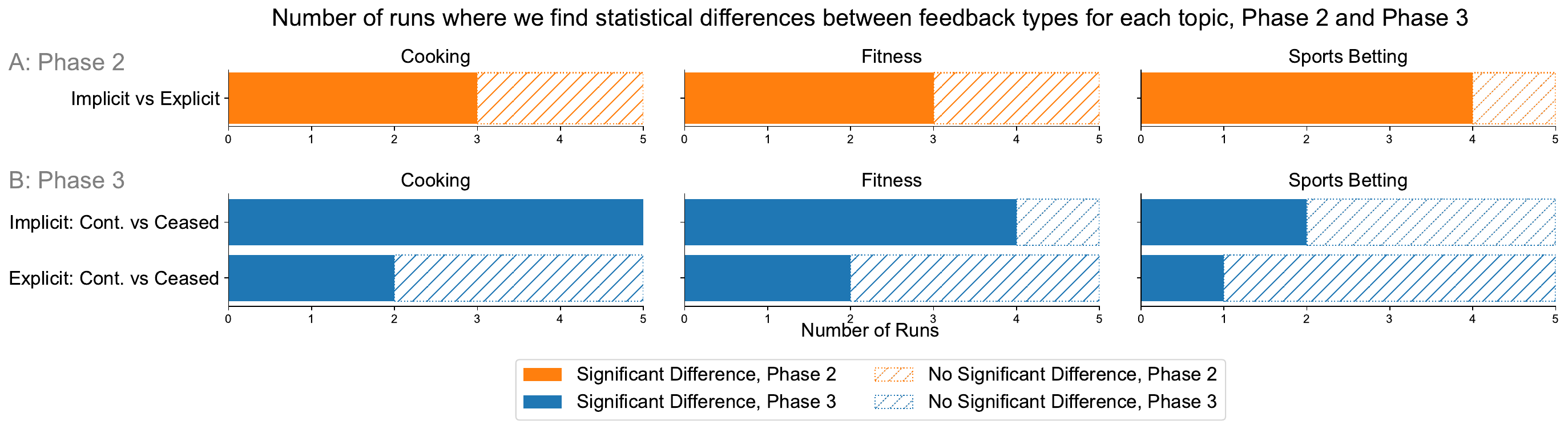}
    \caption{Counts where we find statistical differences for the fifteen experiments across three topics, for Phases 2 and 3. Phase~2~(row \textsf{A}): we typically see statistically more topic videos when implicitly signaling than when explicitly signaling across our three topics. Phase 3 (row \textsf{B}): The \textsf{cooking} and \textsf{fitness} topics consistently relapsed in the implicit case, and relapsed in the explicit case twice. The \textsf{sports betting} topic was found to relapse less often in both the implicit and explicit cases.}
    \label{fig:phase2-phase3-results}
\end{figure*}

The efficacy of the interventions differed, however.
On-topic videos constituted 16\% of the implicit signaling account's feed, compared 30.5\% of the feed of the account which continued watching them (a 47.5\% decrease).
Explicit signaling led to an even more pronounced decrease at 4.75\% (a total decrease of 84.4\%).
In fact, in this experiment, the difference between the prevalence of the on-topic videos in the feed of the explicit signaling account is not significantly different from that of a non-personalized baseline account.

Having shown that both explicit and implicit signaling leads to measurable effects on one topic (cooking), we now turn to including all three target topics and running Phases 2 and 3 of the experiment presented in Figure~\ref{fig:ua-flowchart} five times per topic, for a total of fifteen experimental runs. The number of topic videos delivered to each of our accounts during Phase 2, across all fifteen runs, is shown to the left side of the dotted red line for each of the sub-figures in Figure~\ref{fig:all-fifteen-runs}. Overall we see a reduction in the amount of personalized content over time throughout each of our runs for both implicit and explicit signaling, suggesting that both strategies reduce the amount of topic content on the FYP over time. 

Whether explicit signaling is more effective than implicit signaling differs across topics and runs, however. Figure~\ref{fig:phase2-phase3-results}A tallies the number of runs where a statistical difference was found between the accounts using implicit versus explicit signaling. For each experiment, since we have two accounts for each signaling type, we perform a similar process as before, adding together the total number of topic videos seen for each identical signaling account as the sample proportion, adding up the total number of videos seen (200 per account, for a total of 400) as the sample size, and then running a two-proportion Z-test at the 99\% confidence level. For most runs, we see statistically fewer topic videos for the \textsf{explicit signaling} case compared to  \textsf{implicit signaling}. However, this is not the case for all runs. Sometimes, we see implicit signaling to be as effective as explicit signaling, considerably reducing the amount of topic content seen by the account; this happens for two out of five runs for \textsf{cooking} and \textsf{fitness}, and one out of five runs for \textsf{sports betting}.

\subsection{Phase 3: Treatment Relapse}
Finally, recall that in Phase 3 our goal is to understand to what extent the FYP algorithm starts showing content that a user previously signaled disinterest in, after the signaling ceases. Also recall that Phase 3 introduces a change in behavior for one of the two treatment accounts for each treatment condition. For each target topic and type of signaling, one account continues the signaling, while the other switches to watching on-topic videos. We will refer to the accounts that switch behavior as \textsf{ceases implicit (explicit) signaling} and the accounts that continue signaling as \textsf{continues implicit (explicit) signaling}. The number of topic videos delivered to each of our accounts during Phase 3 can be found to the right side of the dotted red line in Figure~\ref{fig:all-fifteen-runs}.

We define a `relapse' as an instance where the \textsf{ceases implicit (explicit) signaling} account sees statistically more topic videos than the corresponding \textsf{continues implicit (explicit) signaling}. We calculate statistical difference using a two-proportion Z-test at the 99\% confidence level. Recall that we run these accounts simultaneously---we can compare the behavior of the account that continues its negative signaling to the corresponding account that ceases that signaling and begins watching all topic videos to understand whether the algorithm begins pushing more topic content to that account relative to the account that continues negative signaling. 

We run this experiment five times for each topic, continuing the fifteen experiments described in the section above, and measure the number of successful relapses for the implicit signaling and explicit signaling cases. These results can be found in Figure~\ref{fig:phase2-phase3-results}B. We detail the results for \textsf{ceasing implicit signaling} and \textsf{ceasing explicit signaling} below.

\subsubsection{Ceasing Implicit Signaling}
We will first explore the results for relapsing in the \textsf{implicit signaling} case.  For all three topics, we were able to achieve a \textit{relapse} effect multiple times. For the \textsf{cooking} topic, we triggered a relapse for all five runs. This indicates that implicit signaling was not enough to prevent a relapse for an account that returns to watching that content after negatively signaling implicitly for 200 videos.

For \textsf{fitness} and \textsf{sports betting}, we had runs that did not see an implicit relapse. For \textsf{fitness} this only occurred once, which indicates that implicit negative feedback is likely not sufficient in the majority of cases to avoid seeing that content in the FYP in the future. However, we only saw a relapse in two out of five experiments for \textsf{sports betting}. This indicates that relapsing for that topic is less common. Still, with how sensitive these two topics can be, it is interesting that implicit feedback was not enough in multiple cases to prevent the algorithm from pushing that content again.

\subsubsection{Ceasing Explicit Signaling}
We now turn to the accounts that cease explicit signaling (i.e., marking videos `Not Interested'), to investigate how common relapsing is for a stronger negative signal.  We observe fewer relapses compared to the implicit case; two accounts relapsed for the \textsf{cooking} and \textsf{fitness} topics, and one relapsed for the \textsf{sports betting} topic. This indicates that in our controlled environment explicit signaling is more effective than implicit signaling to reduce future exposure should a user's behavior change, across all topics. However, the fact that we were able to trigger a relapse for all three topics indicates that, for some users, their explicit feedback could be overridden if there is a change in behavior in the future.

\section{Discussion}\label{sec:discussion}

In this work, we examined user agency in the context of the FYP algorithm on TikTok, aiming to understand how explicit and implicit signals can be used to control the content users see.
We introduced new methodologies to study user agency in the context of three topics: \textsf{cooking}, \textsf{fitness}, and \textsf{sports betting}. 
We found that, in the experimental setup we created, both explicit signals (i.e., marking a video as `Not Interested') and implicit signals (i.e., skipping past videos) are effective in reducing the delivery of that content. Overall, we find explicit signaling to often be more effective when compared to implicit signaling. However, as we explain below, this finding does not necessarily invalidate the reports of real users who experienced inefficacy of these signals. 

Unfortunately, when investigating a `relapse' effect of switching behavior back to watching that content, we find that accounts that used implicit signals were more often able to `relapse' once such signals ceased. Even when investigating stronger explicit signals, we were still able to trigger a relapse effect for two accounts in the \textsf{cooking} and \textsf{fitness} topics, and one for the \textsf{sports betting} topic.

\subsection{Limitations}
\label{sec:limitations}
Relying on machine-controlled sock-puppet accounts allows for precise isolating of different factors that might contribute to the overall user experience on the platform. 
At the same time, this strength is also the core weakness of the approach.
The effects we observe in this work are strictly driven by the limited set of behaviors we simulated: watching videos, skipping videos, and marking videos as `Not Interested'.
Real users might perform many actions we chose not to simulate: liking, sharing, and reporting videos, commenting on them, as well as following and un-following creators.
Real users' location might further influence their experience, as can their browsing histories, or being targeted by ads.
When real users make a decision on how to react to each video, the signals they consider are different than those available to our sock puppet accounts. 
Additionally, real users might not be focused on one topic exclusively, or as persistent as our sock-puppet accounts in expressing disinterest. 

We chose to focus on a limited number of topics, but there are a wide array of potential interests and real users likely have a wide variety of personalized videos delivered to their feed.
In order to isolate the effect of negative signals our sock-puppet accounts did not substitute the topic of disinterest for a new topic of interest. Real users might, in the process of expressing disinterest in a particular topic, be delivered content that they like and start watching it, thereby shaping their feed more positively.

Finally, we are studying the state of the algorithm at a single point in time. TikTok may adjust how effective negative feedback controls are in controling a user's FYP.  We nevertheless believe that illustrating how effective these controls are in a controlled setting can help users understand how their actions can impact their experience on the platform.

Because of these limitations we do not claim that real users will experience the same extent of personalization in their feed, equivalent levels of efficacy of user controls, or the rate of ``re-personalization'' after they cease to use those controls.
Our results do not undermine the lived experience of users who did not observe the effects of sending explicit signals of disinterest. 
In fact, both can be true---the change introduced by expressing disinterest might be statistically significant, while still not matching real users' expectations.  We hope that, despite its limitations, our research will inform future work with real users.

\subsection{Ethical Considerations}\label{subsec:ethics}
Over the course of this research, we made efforts to ensure ethical data collection and processing. We only collected data from public profiles delivered to the FYP or search feeds, and we did not collect any private user data. We minimized our load on TikTok's servers by only creating the accounts necessary to conduct our experiments. We minimized harm to other TikTok users by only interacting with the platform through watching videos; we did not like videos, comment, or follow creators, to minimize the effect our accounts had on the platform's algorithms. We also avoided watching content promoting illegal or otherwise unethical behavior, choosing topics that represented potentially sensitive but nonetheless allowed content on the platform.

\section*{Acknowledgments}
This work has been funded in part by the National Science Foundation grant CNS-2318290. We thank the anonymous ICWSM Reviewers for their helpful feedback.

\bibliography{references}

\appendix
\renewcommand\thefigure{A\arabic{figure}}
\renewcommand\thetable{A\arabic{table}}
\setcounter{figure}{0}    
\setcounter{table}{0}  

\section*{Ethics Checklist}

\begin{enumerate}

\item For most authors...
\begin{enumerate}
    \item  Would answering this research question advance science without violating social contracts, such as violating privacy norms, perpetuating unfair profiling, exacerbating the socio-economic divide, or implying disrespect to societies or cultures?
    \textcolor{blue}{Yes, this research question seeks to better understand the recommendation algorithm and user agency for users on TikTok}
  \item Do your main claims in the abstract and introduction accurately reflect the paper's contributions and scope?
    \textcolor{blue}{Yes. See Sections~\ref{sec:methodology} and \ref{sec:results}}
   \item Do you clarify how the proposed methodological approach is appropriate for the claims made? 
    \textcolor{blue}{Yes, see section \ref{subsec:design} for a description of how our methodology answers our research questions}
   \item Do you clarify what are possible artifacts in the data used, given population-specific distributions?
    \textcolor{gray}{N/A}
  \item Did you describe the limitations of your work?
    \textcolor{blue}{Yes, the limitations to our methodology are discussed in Section~\ref{sec:limitations}.}
  \item Did you discuss any potential negative societal impacts of your work?
    \textcolor{blue}{We discuss the potential harms of our research methodology in Section~\ref{subsec:ethics}, and the approach we took to mitigating and minimizing those harms.}
      \item Did you discuss any potential misuse of your work?
    \textcolor{blue}{We do not anticipate any misuse, but we do provide discussion on the inaccuracies of applying these findings outside of their scope in Section \ref{sec:limitations}}
    \item Did you describe steps taken to prevent or mitigate potential negative outcomes of the research, such as data and model documentation, data anonymization, responsible release, access control, and the reproducibility of findings?
    \textcolor{blue}{Yes, see Sections~\ref{subsec:emulate} and \ref{sec:feed-cloning} for information on reproducibility and Section~\ref{subsec:ethics} for information on scraping public data.}
  \item Have you read the ethics review guidelines and ensured that your paper conforms to them?
    \textcolor{blue}{Yes.}
\end{enumerate}

\item Additionally, if your study involves hypotheses testing...
\begin{enumerate}
  \item Did you clearly state the assumptions underlying all theoretical results?
    \textcolor{gray}{N/A}
  \item Have you provided justifications for all theoretical results?
    \textcolor{gray}{N/A}
  \item Did you discuss competing hypotheses or theories that might challenge or complement your theoretical results?
    \textcolor{gray}{N/A}
  \item Have you considered alternative mechanisms or explanations that might account for the same outcomes observed in your study?
    \textcolor{gray}{N/A}
  \item Did you address potential biases or limitations in your theoretical framework?
    \textcolor{gray}{N/A}
  \item Have you related your theoretical results to the existing literature in social science?
  \textcolor{gray}{N/A}
  \item Did you discuss the implications of your theoretical results for policy, practice, or further research in the social science domain?
   \textcolor{gray}{N/A}
\end{enumerate}

\item Additionally, if you are including theoretical proofs...
\begin{enumerate}
  \item Did you state the full set of assumptions of all theoretical results?
    \textcolor{gray}{N/A}
	\item Did you include complete proofs of all theoretical results?
    \textcolor{gray}{N/A}
\end{enumerate}

\item Additionally, if you ran machine learning experiments...
\begin{enumerate}
  \item Did you include the code, data, and instructions needed to reproduce the main experimental results (either in the supplemental material or as a URL)?
    \textcolor{blue}{While the main experimental result was not ML focused, we did use a ChatGPT as part of our methodology to classify videos.  Information on our methodology and reproducibility can be found in Sections \ref{sec:methodology} and \ref{sec:results}. We validated our methodology in Section Section~\ref{subsubsec:classify-videos}, and provide inter-rater agreements using Fleiss' Kappa for the four human annotators and the accuracy, precision, recall, and F1 scores for ChatGPT measured against human classification majority vote in Table~\ref{tab:verify-chatgpt-scores}.
    Section~\ref{subsec:iterrater} in the Appendix includes the prompts we gave the LLM. Section~\ref{subsubsec:classify-videos} discusses the reasoning behind using an LLM and why we feel confident in that decision.}
  \item Did you specify all the training details (e.g., data splits, hyperparameters, how they were chosen)?
    \textcolor{gray}{N/A}
     \item Did you report error bars (e.g., with respect to the random seed after running experiments multiple times)?
    \textcolor{gray}{N/A}
	\item Did you include the total amount of compute and the type of resources used (e.g., type of GPUs, internal cluster, or cloud provider)?
    \textcolor{gray}{N/A}
     \item Do you justify how the proposed evaluation is sufficient and appropriate to the claims made? 
    \textcolor{gray}{N/A}
     \item Do you discuss what is ``the cost`` of misclassification and fault (in)tolerance?
    \textcolor{gray}{N/A}
  
\end{enumerate}

\item Additionally, if you are using existing assets (e.g., code, data, models) or curating/releasing new assets, \textbf{without compromising anonymity}...
\begin{enumerate}
  \item If your work uses existing assets, did you cite the creators?
    \textcolor{blue}{Yes, we use multiple different tools throughout the experiment design that is cited in Sections \ref{subsec:emulate}, \ref{sec:watch_decision}, and \ref{sec:feed-cloning}.}
  \item Did you mention the license of the assets?
    \textcolor{blue}{The assets we use are open-source tools or libraries that do not require licenses. }
  \item Did you include any new assets in the supplemental material or as a URL?
    \textcolor{blue}{No.}
  \item Did you discuss whether and how consent was obtained from people whose data you're using/curating?
    \textcolor{blue}{Yes, see Section~\ref{subsec:ethics}.}
  \item Did you discuss whether the data you are using/curating contains personally identifiable information or offensive content?
    \textcolor{blue}{We only collect data that has been posted publicly, and we do not collect private or personally identifiable information about any users or creators on the platform. We avoid watching content that promotes illegal/unethical behavior, and only engage with specific categories of content that are allowed on TikTok. The sensitive topics we look at in this paper may be considered offensive, as discussed in Sections \ref{subsec:fitness} and \ref{subsec:sportsbetting}, however they do not violate any community or content guidelines. This is further discussed in Section \ref{subsec:ethics}.}
\item If you are curating or releasing new datasets, did you discuss how you intend to make your datasets FAIR (see \citet{fair})?
\textcolor{blue}{No new datasets are being curated or released, }
\item If you are curating or releasing new datasets, did you create a Datasheet for the Dataset (see \citet{gebru2021datasheets})? 
\textcolor{gray}{N/A}
\end{enumerate}

\item Additionally, if you used crowdsourcing or conducted research with human subjects, \textbf{without compromising anonymity}...
\begin{enumerate}
  \item Did you include the full text of instructions given to participants and screenshots?
    \textcolor{gray}{N/A}
  \item Did you describe any potential participant risks, with mentions of Institutional Review Board (IRB) approvals?
    \textcolor{gray}{N/A}
  \item Did you include the estimated hourly wage paid to participants and the total amount spent on participant compensation?
    \textcolor{gray}{N/A}
   \item Did you discuss how data is stored, shared, and deidentified?
   \textcolor{gray}{N/A}
\end{enumerate}

\end{enumerate}

\section{Appendices}

\subsection{ChatGPT Prompts}\label{subsec:iterrater}
We used the following prompt structure, with the target topic and keywords presented in Table~\ref{tab:verify-chatgpt-scores} substituted in: 

``You are a classifier tasked with determining whether the given content has anything to do with [keywords]. Given a list of: the user who posted a video and a brief description of them; the video's description; a list of related words; and a list of hashtags, you classify whether the video is related to [topic]. You only respond with `Yes' if you think it is, or `No' if not.

Description: [...], Hashtags: [...], Suggested Words: [...], Nickname: [...], Signature: [...]''

\subsection{Dates for Experiments}
Table~\ref{tab:dates-table} presents the dates when each Phase of each of our experiments were run. We ran Phase 1 once for each topic. We ran Phases 2 and 3 five times for each topic, for a total of 15 runs. 

\begin{table*}[h!]
 \setlength{\tabcolsep}{2mm}
\begin{tabular}{@{}l|l|llllllllll@{}}
\toprule
\multicolumn{1}{c|}{} & \textbf{Phase 1} & \multicolumn{10}{c}{\textbf{Phases 2/3}}                                                                                                                                                      \\ \cmidrule(l){2-12} 
Topic                 &                  & \multicolumn{2}{c|}{Run 1}             & \multicolumn{2}{c|}{Run 2}             & \multicolumn{2}{c|}{Run 3}             & \multicolumn{2}{c|}{Run 4}             & \multicolumn{2}{c}{Run 5} \\ \cmidrule(l){3-12} 
                      &                  & Phase 2 & \multicolumn{1}{l|}{Phase 3} & Phase 2 & \multicolumn{1}{l|}{Phase 3} & Phase 2 & \multicolumn{1}{l|}{Phase 3} & Phase 2 & \multicolumn{1}{l|}{Phase 3} & Phase 2     & Phase 3     \\ \midrule
Cooking               & 3/17/25          & 3/27/25 & \multicolumn{1}{l|}{3/27/25} & 3/31/25 & \multicolumn{1}{l|}{3/31/25} & 4/11/25 & \multicolumn{1}{l|}{4/11/25} & 4/15/25 & \multicolumn{1}{l|}{4/15/25} & 4/17/25     & 4/17/25     \\
Fitness               & 3/17/25          & 4/4/25  & \multicolumn{1}{l|}{4/4/25}  & 4/9/25  & \multicolumn{1}{l|}{4/9/25}  & 4/16/25 & \multicolumn{1}{l|}{4/16/25} & 4/21/25 & \multicolumn{1}{l|}{4/21/25} & 4/22/25     & 4/22/25     \\
Betting        & 3/17/25          & 4/10/25 & \multicolumn{1}{l|}{4/10/25} & 4/10/25 & \multicolumn{1}{l|}{4/10/25} & 4/14/25 & \multicolumn{1}{l|}{4/14/25} & 4/16/25 & \multicolumn{1}{l|}{4/16/25} & 4/18/25     & 4/18/25     \\ \bottomrule
\end{tabular}
\caption{The dates when our various experiments were run.}
\label{tab:dates-table}
\end{table*}

\end{document}